\documentclass[prd,preprint,showpacs,groupedaddress]{revtex4-1}
\usepackage{amsmath}
\usepackage{amsfonts}
\usepackage{amssymb}
\usepackage{geometry}
\usepackage{natbib}
\usepackage[english]{babel}
\usepackage{CJK}
\usepackage{graphicx}
\usepackage{fancyhdr}
 \usepackage{epstopdf}
\begin{document}

  \title{Thermodynamics of charged AdS black holes in rainbow gravity}

  \author{Ping Li} \author{Miao He} \author{Jia-Cheng Ding} \author{Xian-Ru Hu} \author{Jian-Bo Deng}\email[Jian-Bo Deng:]{dengjb@lzu.edu.cn}

  \affiliation{Institute of Theoretical Physics, LanZhou University,
    Lanzhou 730000, P. R. China}

  \date{\today}
\begin{abstract}
In this paper, the thermodynamic property of charged AdS black holes is studied in rainbow gravity. By the Heisenberg Uncertainty Principle and the modified dispersion relation, we obtain deformed temperature. Moreover, in rainbow gravity we calculate the heat capacity in a fixed charge and discuss the thermal stability. We also obtain a similar behaviour with the liquid-gas system in extending phase space (including \(P\) and \(r\)) and study its critical behavior with the pressure given by the cosmological constant and with a fixed black hole charge \(Q\). Furthermore, we study the Gibbs function and find its characteristic swallow tail behavior, which indicates the phase transition. We also find there is a special value about the mass of test particle which would lead the black hole to zero temperature and a diverging heat capacity with a fixed charge.

\end{abstract}

  \pacs{04.70.-s, 04.50.Kd}

  \keywords {energy-momentum dispersion relation, doubly special relativity,thermodynamics and phase transition,Charged AdS black holes}

  \maketitle

  \section{Introduction}
 It is known Lorentz symmetry is one of most important symmetries in nature, however, some researches indicate the Lorentz symmetry might be violated in the ultraviolet limit~\cite{iengo2009renormalization,adams2006causality,gripaios2004modified,alfaro2015electroweak,belich2014geometric}. Since the standard energy-momentum dispersion relation relates to the Lorentz symmetry, the deformation of Lorentz symmetry would lead to the modification of energy-momentum dispersion relation. In fact, some calculations in loop quantum gravity have showed the dispersion relations may be deformed. Meanwhile, based on the deformed energy-momentum dispersion relation the double special relativity has arisen~\cite{PhysRevD.71.026010,PhysRevD.67.044017}. In this theory, in addition to the velocity of light being the maximum velocity attainable there is another constant for maximum energy scale in nature which is the Planck energy \(E_{P}\). It gives different picture for the special relativity in microcosmic physics. The theory has been generalized to curved spacetime by Joao Magueijo and Lee Smolin, called gravity's rainbow~\cite{magueijo2004gravity}. In their theory, the geometry of spacetime depends on the energy of the test particle and observers of different energy would see different geometry of spacetime. Hence, a family of energy-dependent metrics named as rainbow metrics will describe the geometry of spacetime, which is different from general gravity theory. Based on the non-linear of Lorentz transformation, the energy-momentum dispersion relation can be rewritten as
\begin{equation}\label{one}
  E^{2}f^{2}(E/E_{P})-p^{2}g^{2}(E/E_{P})=m^{2},
\end{equation}
where \(E_{P}\) is the Planck energy. The rainbow functions \(f(E/E_{P})\) and \(g(E/E_{P})\) are required to satisfy
\begin{equation}
  \lim_{E/E_{P}\to 0}f(E/E_{P})=1,~~\lim_{E/E_{P}\to 0}g(E/E_{P})=1.
\end{equation}
In this case, the deformed energy-momentum dispersion relation Eq.(\ref{one}) will go back to classical one when the energy of the test particle is much lower than \(E_p\). Due to this energy-dependent modification to the dispersion relation, the metric \(h(E)\) in gravity's rainbow could be rewritten as~\cite{peng2008covariant}
\begin{equation}
  h(E)=\eta^{ab} e_{a}(E)\otimes e_{b}(E),
\end{equation}
where the energy dependence of the frame fields are
\begin{equation}
  e_{0}(E)=\frac{1}{f(E/E_{P})} \widetilde{e_{0}}~,~~ e_{i}(E)=\frac{1}{g(E/E_{P})} \widetilde{e_{i}},
\end{equation}
here the tilde quantities refer to the energy-independent frame fields. This leads to a one-parameter Einstein equation
 \begin{equation}\label{1}
  G_{\mu\nu}(E/E_{P})+\Lambda(E/E_{P})g_{\mu\nu}(E/E_{P})=8\pi G(E/E_{P})T_{\mu\nu}(E/E_{P}),
\end{equation}
where \(G_{\mu\nu}(E/E_{P})\) and \(T_{\mu\nu}(E/E_{P})\) are energy-dependent Einstein tensor and energy-momentum tensor, \(\Lambda(E/E_{P})\) and \(G(E/E_{P})\) are energy-dependent cosmological constant and Newton constant. Generally, many forms of rainbow functions have been discussed in literatures, in this paper we will mainly employ the following rainbow functions
\begin{equation}\label{11}
  f(E/E_{P})=1,~~g(E/E_{P})=\sqrt{1-\eta (E/E_{P})^{n}} ,
\end{equation}
which has been widely used in Refs~\cite{amelino1997distance,jacob2010modifications,amelino2013quantum,ali2014black,
ali2015absence,ali2015remnant,hendi2015black,hendi2016charged,gangopadhyay2016constraints}.

 Recently, Schwarzschild black holes, Schwarzschild AdS black holes and Reissner-Nordstrom black holes in rainbow gravity~\cite{gim2014thermodynamic,kim2016thermodynamic,liu2014charged} have been studied.
 Ahmed Farag Alia, Mir Faizald and Mohammed M.Khalile~\cite{ali2015remnant} studied the deformed temperature about charged AdS black holes in rainbow gravity based on Heisenberg Uncertainty Principle(HUP), $E=\Delta p\sim\frac{1}{r_{+}}$. In this paper, we study the thermodynamical property about the charged AdS black holes in rainbow gravity based on the usual HUP, \( p=\Delta p\sim\frac{1}{r_{+}}\). Moreover, we study how the mass of test particle influences thermodynamical property for charged AdS black holes.

 The paper is organized as follows. In the next section, by using the HUP and the modified dispersion relation, we obtain deformed temperature, we also calculate heat capacity with a fixed charge and discuss the thermal stability.
  In Sec.\uppercase\expandafter{\romannumeral3}, we find the charged AdS black holes have similar behaviour with the liquid-gas system with the pressure given by the cosmological constant while we treat the black holes charge \(Q\) as a fixed external parameter, not a thermodynamic variable. We also calculate the Gibbs free energy and find characteristic swallow tail behavior.  Finally, the conclusion and discussion will be offered in Sec.\uppercase\expandafter{\romannumeral4}.

\section{The thermal stability}
In rainbow gravity the line element of the modified charged AdS black holes can be described as~\cite{ali2015remnant}
\begin{equation}
  ds^{2}=-\frac{N}{f^{2}}dt^{2}+\frac{1}{Ng^{2}}dr^{2}+\frac{r^{2}}{g^{2}}d\Omega^{2},
\end{equation}
where
\begin{equation}\label{121}
 N=1-\frac{2M}{r}+\frac{Q^{2}}{r^{2}}+\frac{r^{2}}{l^{2}}.
\end{equation}
 Generally, \(-\frac{3}{l^{2}}=\Lambda\) which is cosmological constant. Because all energy dependence in the energy-independent coordinates must be in the rainbow functions \(f\) and \(g\), \(N\) is independent on the energy of test particle~\cite{magueijo2004gravity}. In gravity's rainbow, the deformed temperature related to the standard temperature \(T_{0}\) was~\cite{ali2015remnant}
 \begin{equation}\label{18}
   T=-\frac{1}{4\pi} \lim_{r \to r_{+}} \sqrt{\frac{-g^{11}}{g^{00}}} \frac{(g^{00})^{'}}{g^{00}}=\frac{g(E/E_{P})}{f(E/E_{P})} T_{0},
 \end{equation}
where \(r_+\) is horizon radius.

 In gravity's rainbow, although the metric depends on the energy of test particle, the usual HUP can be still used~\cite{gim2014thermodynamic}. For simplicity we take $n=2$ in the following discussion, by combining Eq.(\ref{one}) with Eq.(\ref{11}) we can get
 \begin{equation}\label{3}
    g=\sqrt{1-\eta G_{0} m^{2}} \sqrt{\frac{r_{+}^{2}}{r_{+}^{2}+\eta G_{0}}},
 \end{equation}
 where \(G_{0}=1/E_{P}^2\), \(m\) is the mass of test particle and \(\eta\) is a constant parameter.

Generally, the standard temperature was given by~\cite{kubizvnak2012p}
\begin{equation}
  T_0=\frac{1}{4\pi}(\frac{1}{r_{+}}+\frac{3r_{+}}{l^{2}}-\frac{Q^{2}}{r_{+}^{3}}).
\end{equation}
When using Eqs.(\ref{11}) and (\ref{3}), we can get the temperature of charged AdS black holes in rainbow gravity
 \begin{equation}\label{4}
   T=gT_{0}=\frac{1}{4\pi k} \sqrt{\frac{r_{+}^{2}}{r_{+}^{2}+\eta G_{0}}}  (\frac{1}{r_{+}}+\frac{3r_{+}}{l^{2}}-\frac{Q^{2}}{r_{+}^{3}}),
 \end{equation}
 where \(k=1/ \sqrt{1-\eta G_{0} m^{2}}\). It is easy to find \(T=T_0\) when \(\eta=0\). Eq.(\ref{4}) shows there are two solutions when \(T=0\), one corresponds to extreme black hole, the other to \(m^{2}=\frac{1}{\eta G}\). The second solution indicates the temperature of black holes completely depends on the mass of test particle when the black holes keep with fixed mass, charge and anti-de Sitter radius. The bigger the mass of test particle is, the smaller the temperature of black holes is. When \(m^{2}=\frac{1}{\eta G}\), the temperature keeps zero. Generally, due to gravity's rainbow, a minimum radius with respect to the black hole is given and is related to a radius of black hole remnant when the temperature tends to zero ~\cite{ali2015remnant}. However, our paper shows all black holes can keep zero temperature when the test particle mass approaches a value, such as, \(m^{2}=\frac{1}{\eta G}\). But due to \(m\ll M_{P}\) in general condition, it may be difficult to test the phenomenon with zero temperature about black holes.

  \begin{figure}[htbp]
     \centering\includegraphics[width=10cm]{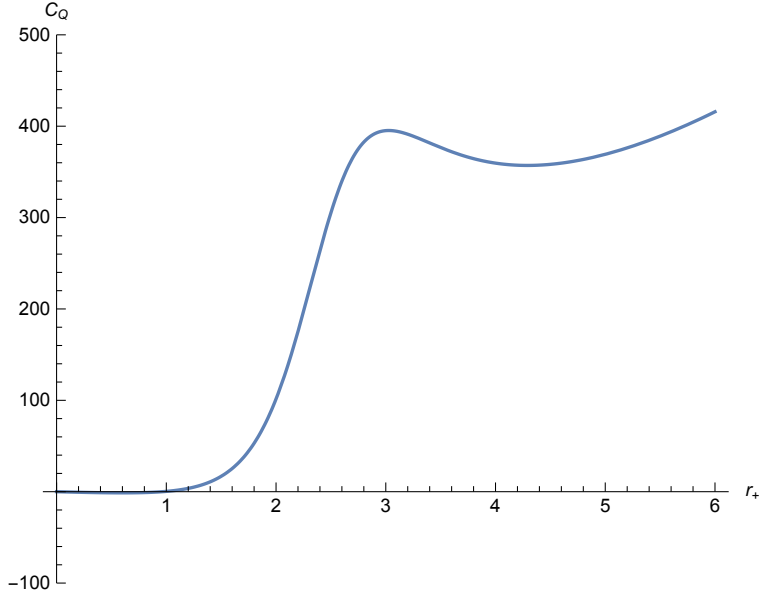}
   \caption{\(C_Q -r_{+}\) diagram of charged AdS black holes in the rainbow gravity. It corresponds to \(l=6\). We have set \(Q=1,\eta =1,m=0\).}\label{fig:04}
\end{figure}
\begin{figure}[htbp]
    \centering\includegraphics[width=10cm]{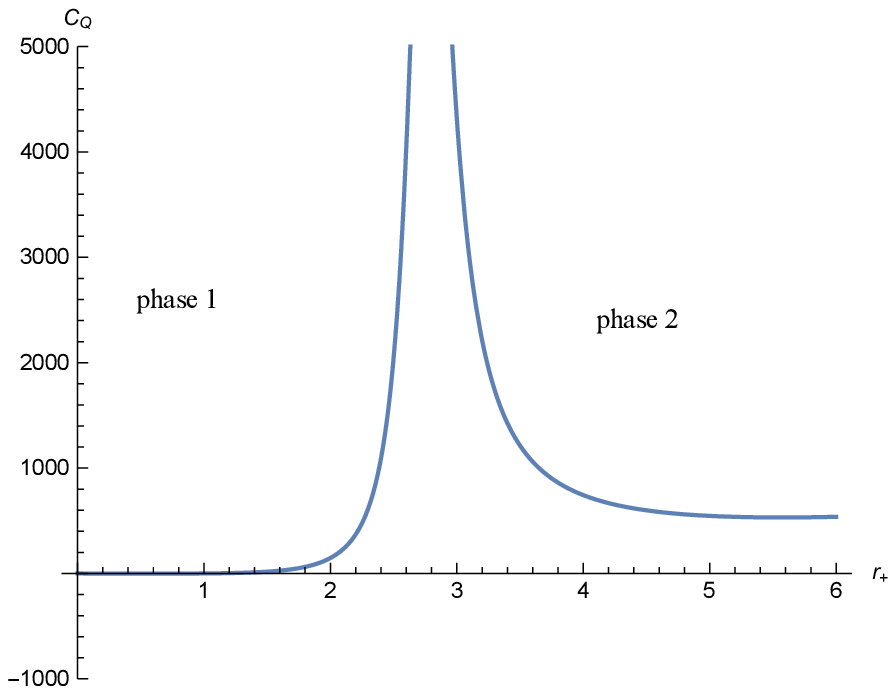}
    \caption{\(C_Q-r_{+}\) diagram of charged AdS black holes in the rainbow gravity. It corresponds to \(l=7.05\). We have set \(Q=1,\eta =1,m=0\).}\label{fig:02}
\end{figure}
 \begin{figure}[htbp]
   \centering\includegraphics[width=5cm]{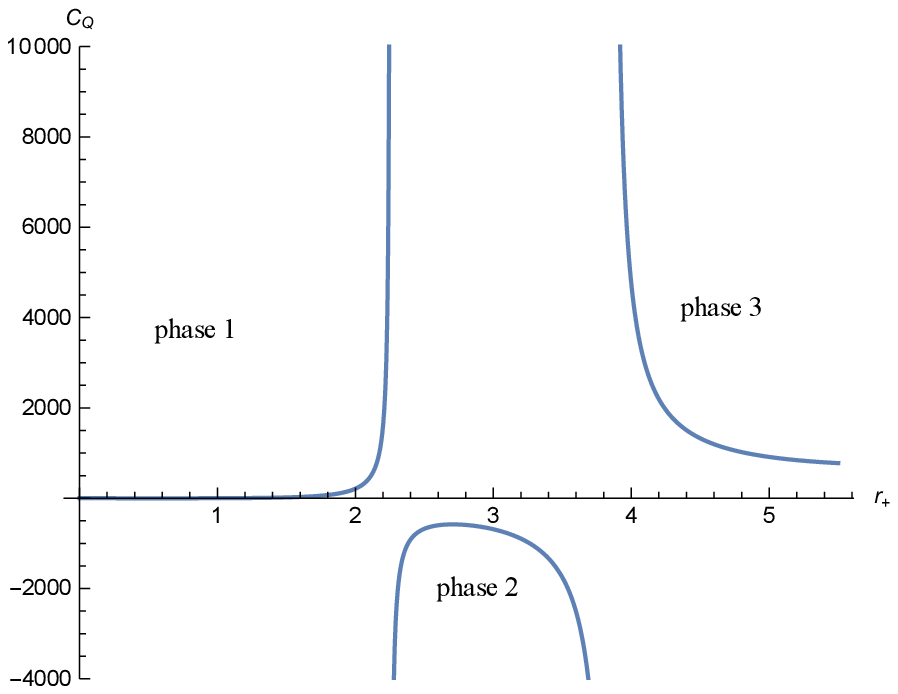}
   \includegraphics[width=5cm]{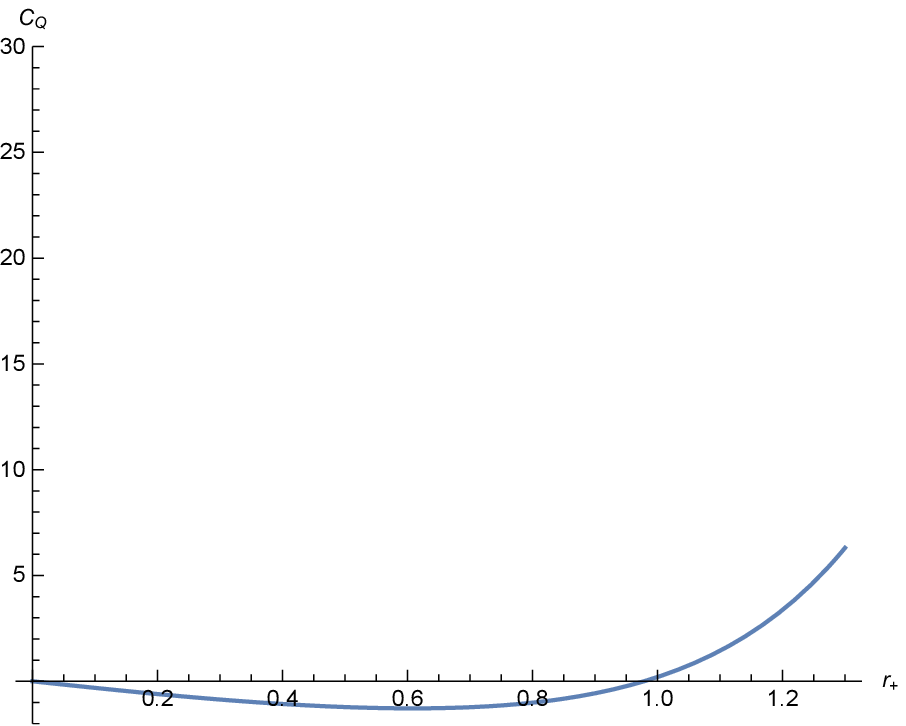}
   \caption{\(C_Q -r_{+}\) diagram of charged AdS black holes in the rainbow gravity. It corresponds to \(l=8\), right one corresponds to an extending part of near \(r_+=1\). We have set \(\eta =1,Q=1,m=0\).}\label{fig:03}
\end{figure}

  In general, the thermal stability can be determined by the heat capacity, which is also used to the systems of black holes~\cite{hendi2016charged,1,2,4}. In other words, the positive heat capacity corresponds to a stable state and the negative heat capacity corresponds to unstable state. In following discussions, we will focus on the heat capacity to discuss the stability of black holes. When \(N=0\), the mass of charged AdS black holes can be calculated as
 \begin{equation}
   M=\frac{1}{2} (r_{+}+\frac{Q^{2}}{r_{+}}+\frac{r_{+}^{3}}{l^{2}}).
 \end{equation}
   Based on the first law \(dM=TdS\) with the deformed temperature\cite{kim2016thermodynamic}, the modified entropy can be computed
  \begin{equation}
 \begin{aligned}
  S=&\int \frac{dM}{T}\\
  =&\pi k r_{+}\sqrt{r_{+}^{2}+\eta G_{0}}+\pi k \eta G_{0}\ln(r_{+}+\sqrt{r_{+}^{2}+\eta G_{0}} ).
   \end{aligned}
\end{equation}
 Note that the next leading order is logarithmic as \(S\approx \pi r_+^{2}+\frac{1}{2} \pi \eta G_0 \ln(4r_{+}^{2})\), which is similar to the quantum correction in Refs.\cite{fursaev1995temperature,kaul2000logarithmic,das2002general,chatterjee2004universal,wang2008entropy,3}. With \(A=4\pi r_+^{2}\) we can get \(S\approx \frac{A}{4}+\frac{1}{2} \pi \eta G_0 \ln(\frac{A}{\pi})\). We can find the result is in agreement with the standard entropy \(S=A/4\) when \(\eta =0\), which is standard condition.

The heat capacity with a fixed charge can be calculated as
\begin{equation}\label{8}
\begin{aligned}
 C_{Q}&=T\frac{dS}{dT}=(\frac{{\partial M }/{\partial r_{+}}}{{\partial T }/{\partial r_{+}}})\\
 &=2\pi k \frac{(-Q^{2}l^{2}r_{+}^{2}+l^{2}r_{+}^{4}+3r_{+}^{6})(r_{+}^{2}+\eta G_{0})^{3/2}}{3r_{+}^{7}+(6\eta G_{0}-l^{2})r_{+}^{5}+3Q^{2}l^{2}r_{+}^{3}+2\eta  G_{0}Q^{2}l^{2}r_{+}},
  \end{aligned}
\end{equation}
 which shows that \(C_Q\) reduces to standard condition~\cite{kubizvnak2012p} with \(\eta=0\). Obviously, the  heat capacity is diverging when \(m^{2}=\frac{1}{\eta G}\). Generally, when the temperature vanishes, the heat capacity also tends to zero. However, our paper shows a different and anomalous phenomenon. Fortunately, the phenomenon is just an observation effect, the result gives us a way to test the theory of rainbow gravity. Some of the conditions above indicate that the mass of test particle does not influence the forms of temperature, entropy and heat capacity but only changes their amplitudes.

  The numerical methods indicate there are three situations corresponding to zero, one and two diverging points of heat capacity respectively, which have been described in Fig.\ref{fig:04}, Fig.\ref{fig:02}, Fig.\ref{fig:03}. Fig.\ref{fig:04} shows a continuous phase and does not appear phase transition with \(l<l_c\). In Fig.\ref{fig:02}, there is one diverging point and two stable phases for \(C_Q>0\) with \(l=l_{c}\), phase 1 and phase 2, which individually represents a phase of large black hole (LBH) and a small black hole (SBH). In Fig.\ref{fig:03}, one can find there are three phases and two diverging points with \(l>l_{c}\). Phase 1 experiences a continuous process from a unstable phase \(C_Q<0\) to a stable phase \(C_Q>0\); phase 2 is a pure unstable phase with \(C_Q<0\); phase 3 is a stable phase with \(C_{Q}>0\). It is easy to see phase 1 represents the phase of SBH and phase 3 represents the phase of LBH. However, there is a special unstable phase 2 between phase 1 and phase 3. This indicates when the system evolutes from phase 3 to phase 1, the system must experience a medium unstable state which could be explained as an exotic quark-gluon plasma with negative heat capacity ~\cite{burikham2014,burikham2016mixed}.
 \section{The phase transition of Charged AdS black holes in extending phase space}
 Surprisingly, although rainbow functions modify the \(\Lambda (E/E_{P})\) term, they do not affect thermodynamical pressure related to the cosmological constant~\cite{hendi2016charged,PhysRevD.94.024028}. So we can take the following relation
 \begin{equation}\label{5}
   P=-\frac{\Lambda (0)}{8\pi}=\frac{3}{8\pi l^{2}}.
 \end{equation}
 \begin{figure}[htbp]
   \centering\includegraphics[width=10cm]{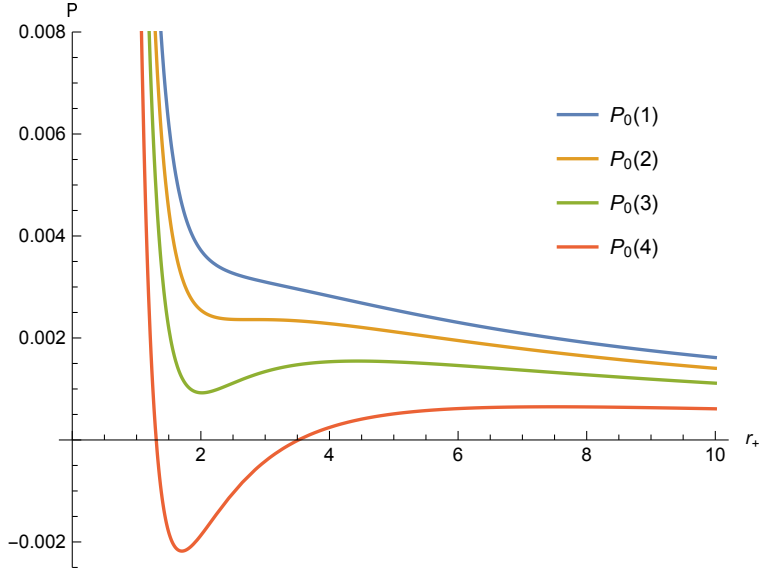}
   \caption{\(P-r_{+}\) diagram of charged AdS black holes in the rainbow gravity. The temperature of isotherms decreases from top to bottom. The \(P_0(1)\) line corresponds to one-phase for \(T>T_{c}\). The critical state, \(T_{c}=0.0358\), is denoted by the \(P_0(2)\) line. The lowest two lines correspond to the smaller temperature than the critical temperature. We have set \(Q=1,\eta =1,m=0\).}\label{fig:1}
\end{figure}

 Since David Kubiznak and Robert B. Mann have showed the critical behaviour of charged AdS black holes and completed the analogy of this system with the liquid-gas system~\cite{kubizvnak2012p}, in what follows we will study whether the critical behavior of the charged AdS black holes system in rainbow gravity is kept. Using Eq.(\ref{4}) and Eq.(\ref{5}) in extended phase space, we can get
 \begin{equation}
   P=\frac{k}{2} \sqrt{\frac{r_{+}^{2}+\eta G_{0}}{r_{+}^{4}}} T-\frac{1}{8\pi} \frac{1}{r_{+}^{2}}+\frac{1}{8\pi} \frac{Q^{2}}{r_{+}^{4}}.
 \end{equation}
 Similarly with Ref.\cite{hendi2016charged}, the critical point is obtained from
\begin{equation}
  \frac{\partial P}{\partial r_+}=0 , ~~~\frac{\partial^2 P}{\partial r_+^2}=0,
\end{equation}
which leads to
\begin{equation}\label{009}
\begin{aligned}
  &r_c=\sqrt{\frac{2^{4/3}\eta G_{0} Q^{2}+2^{4/3}Q^{4}+2Q^{2}(x+y)^{1/3}+2^{2/3}(x+y)^{2/3}}{(x+y)^{1/3}}}    ,\\
      &T_{c}=\frac{1}{2\pi k} \frac{r_{c}^{2}-2Q^2}{r_{c}^{4}+2\eta G_{0}r_{c}^{2}} \sqrt{r_{c}^{2}+\eta G_{0}} , \\
      &P_{c}=\frac{r_c^4-3Q^2r_c^2-2\eta G_0 Q^2}{8 \pi r_c^4(r_c^2+2\eta G_0)},
  \end{aligned}
 \end{equation}
where \(x=\eta G_{0} Q^{2}(\eta G_{0}+Q^{2})\), \(y=Q^{2}(\eta G_{0}+Q^{2})(\eta G_{0}+2Q^{2})\). We can obtain
 \begin{equation}\label{35}
   \frac{P_c r_c}{T_c}=k \frac{r_c^4-3Q^2r_c^2-2\eta G_0 Q^2}{4 r_c(r_c^2-2Q^2)\sqrt{r_c^2+\eta G}},
 \end{equation}
 which shows the critical ratio is deformed due to the existence of rainbow gravity. It is notable that Eq.(\ref{35}) will back to the usual ratio with \(\eta=0\) . Generally, for charged AdS black holes, the pressure and temperature are demanded as positive real value. From Eq.(\ref{009}), when \(P_c>0\) and \(T_c>0\), we have
 \begin{equation}\label{102}
   r_c^4-3Q^2r_c^2-2\eta G_0Q^2>0 ,r_c>\sqrt{2} Q ,
 \end{equation}
 which indicates a restriction between \(Q\) and \(\eta\). The \(P-r_+\) diagram has been described in Fig.\ref{fig:1}. From Fig.\ref{fig:1}, we can find that charged AdS black holes in rainbow gravity have an analogy with the Van-der-Waals system and have a first-order phase transition with \(T<T_c\). Namely, when considering rainbow gravity with the form of Eq.(\ref{11}), the behavior like Van-der-Waals system can also be obtained.
 \begin{figure}[htbp]
   \centering\includegraphics[width=8cm]{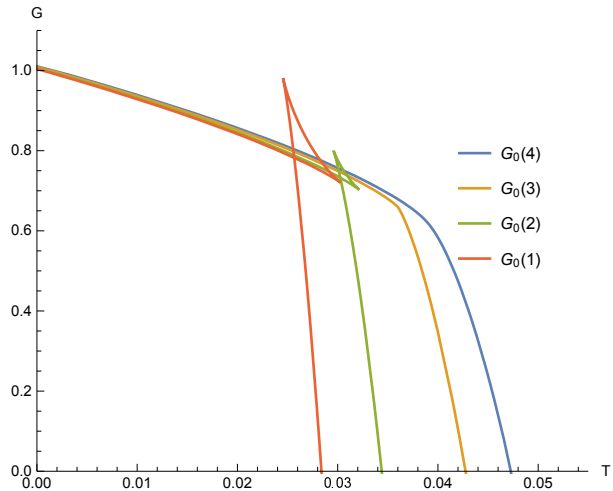}
   \caption{ Gibbs free energy of charged AdS black holes in rainbow gravity. The blue line \(G_0(3)\) corresponds to the critical pressure \(P_{c}\approx 0.0024\), the line \(G_0(4)\) corresponds to pressure \(P>P_{c}\), and the others corresponds to pressure \(P<P_{c}\). We have set \(Q=1,\eta =1,m=0\).}\label{fig:5}
\end{figure}

  Based on Ref.\cite{kastor2009enthalpy,ali2015absence} the black hole mass is identified with the enthalpy, rather than the internal energy, so the Gibbs free energy for fixed charge in the rainbow gravity will be
 \begin{equation}\label{10}
 \begin{aligned}
   G&=H-TS\\
   &=\frac{1}{2} (r_{+}+\frac{Q^{2}}{r_{+}}+\frac{r_{+}^{3}}{l^{2}})\\
   &-k T(\pi r_{+}\sqrt{r_{+}^{2}+\eta G_{0}}+\pi \eta G_{0}\ln(r_{+}+\sqrt{r_{+}^{2}+\eta G_{0}}),
  \end{aligned}
 \end{equation}
  which has been showed in the Fig.\ref{fig:5}. Because the picture of \(G\) demonstrates the characteristic ¡®swallow tail¡¯ behaviour, there is a first order transition in the system.

\section{Conclusion}
 In this paper, we have studied the thermodynamic behavior of charged AdS black holes in rainbow gravity. By the modified dispersion relation and HUP, we got deformed temperature in charged AdS black holes using no-zero mass of test particle. We have discussed the divergence about the heat capacity with a fixed charge. Our result shows that the phase structure has a relationship with AdS radius $l$. When $l=l_c$, there is only one diverging point about heat capacity; when $l>l_c$, we have found there are two diverging points and three phases including two stable phases and one unstable phase. In particular, an analogy between the charged AdS black holes in the rainbow gravity and the liquid-gas system is discussed. We have also showed \(P-r_+\) critical behavior about the charged AdS black holes in the rainbow gravity. The consequence shows there is the Van-der-Waals like behavior in the rainbow gravity when \(\eta\) and \(Q\) coincide with Eq.(\ref{102}). The rainbow functions deform the forms of critical pressure,temperature and radius. At last, we have discussed the Gibbs free energy and have obtained characteristic `swallow tail' behaviour which can be the explanation of first-order phase transition.

We find the mass of test particle does not influence the forms of temperature, entropy and heat capacity but only changes their amplitudes. Moreover, there is a special value about the mass of test particle encountered \(m^{2}=\frac{1}{\eta G}\), which would lead to zero temperature and diverging heat capacity for charged AdS black holes in rainbow gravity.

\section*{Conflicts of Interest}
The authors declare that there are no conflicts of interest regarding the publication of this paper.

  \section*{Acknowledgments}
 We would like to thank the National Natural Science Foundation of China (Grant No.11571342) for supporting us on this work.

  \section*{References}

 \bibliographystyle{unsrt}
 \bibliography{reference}

\end{document}